\documentclass[aps,twocolumn,floatfix, showkeys, superscriptaddress, nofootinbib]{revtex4-1}
\usepackage{amsmath}
\usepackage{graphicx,bm,hhline, xcolor, float}

\begin{document}
\title{Average Atom Model with Siegert States}

\author{C. E. Starrett}
\email{starrett@lanl.gov}
\affiliation{Los Alamos National Laboratory, P.O. Box 1663, Los Alamos, NM 87545, U.S.A.}

\author{N. R. Shaffer}
\affiliation{Los Alamos National Laboratory, P.O. Box 1663, Los Alamos, NM 87545, U.S.A.}
\affiliation{Laboratory for Laser Energetics, University of Rochester, 250 East River Road, Rochester, NY 14623, U.S.A.}

\date{\today}
\begin{abstract}
In plasmas, electronic states can be well localized bound states or itinerant free states, or something in between.  In self-consistent treatments of plasma electronic structure such as the average atom model, all states must be accurately resolved in order to achieve a converged numerical solution.  This is a challenging numerical and algorithmic problem in large part due to the continuum of free states which is relatively expensive and difficult to resolve accurately.  Siegert states are an appealing alternative.  They form a complete eigenbasis with a purely discrete spectrum while still being equivalent to a representation in terms of the usual bound states and free states. However, many of their properties are unintuitive, and it is not obvious that they are suitable for self-consistent plasma electronic structure calculations.  Here it is demonstrated that Siegert states can be used to accurately solve an average atom model and offer advantages over the traditional finite-difference approach, including a concrete physical picture of pressure ionization and continuum resonances.
\end{abstract}
\maketitle

\section{Introduction}
Average atom models use finite temperature density functional theory \cite{mermin65} to calculate the average electronic structure of one atom in a material.  The aim is to have a fast but reasonably accurate physical model of the electronic structure, and yield information like the equation of state and average ionization.  There has been a long standing effort to make the numerical implementation of these models robust yet inexpensive \cite{liberman, more1985pressure, blenski95, wilson06, piron11, starrett19, scaalp}.

%The key numerical challenge is finding all the eigenstates of the Hamiltonian, corresponding to both bound and free states, and accurately integrating over all these states (over energy).  In principle this demands a numerical scheme that can quickly and accurately find all the bound solutions, no matter how weakly bound \cite{vcertik2013dftatom}, as well as the ability to find and resolve resonances in the continuum of free states.
%% Here's my attempt:
The essential step in evaluating an average atom model is the self-consistent solution of the one-electron Schr\"odinger (Kohn-Sham) equation in which the effective potential is a functional of the total electron density.  The electron density contains contributions from both the bound and the free states.  One needs to quickly and accurately search for all the bound states, no matter how weakly bound \cite{vcertik2013dftatom}.  One also needs to solve for a large number of continuum states, especially at conditions where sharp resonances appear in the density of states. These resonances are difficult to track, since they can appear and disappear over the course of a self-consistent iteration \cite{wilson06}. 

Methods of solving the Kohn-Sham equations \cite{kohn1965self} in average atoms typically involve numerical finite-difference integration schemes like Adams-Bashforth or Runge-Kutta.  A largely unexplored alternative is to use basis functions via spectral methods, where one solves a matrix problem for the coefficients.  The primary problem is that the boundary condition for the average atom requires that the wavefunctions match the free electron solutions at the cell boundaries (or beyond, depending on the model).  This condition admits a discrete bound state spectrum and a continuous free electron spectrum.  Thus, to use basis function methods one would need to discretize the continuum, introducing a physical approximation in doing so \cite{peyrusse2006use}.

In general, average atom eigenfunctions are found by matching to the physical free electron solutions, which are combinations of incoming and outgoing waves, i.e., these are the physical scattering solutions that are familiar to us.  An interesting alternative set of eigenstates is found if one retains the free electron boundary condition but restricts the eigenfunction to have outgoing character only.  These states were introduced in 1939 by Siegert, motivated by the search for a formal derivation of the Breit-Wigner formula \cite{siegert1939derivation}.  These Siegert states, as they are now known, encompass the usual discrete bound states, but in place of the real-energy continuum of free states, a discrete set of  complex-energy eigenstates is found.  These Siegert States influence physical quantities but are themselves not directly observable.  However, the observable physical states are uniquely related to the Siegert states via simple summations.

In the decades after Siegert's paper, it proved difficult to find an efficient method for solving for the Siegert states, until a basis function method was ultimately derived \cite{tolstikhin97siegert}.  This method assumes a finite ranged potential ($V(r) = 0$ for $r>a$), and because of this and the finite number of basis terms, the authors referred to their solutions as Siegert Pseudo-States.  The original papers \cite{tolstikhin97siegert, tolstikhin1998siegert} were restricted to s-waves only.  However, the general solution for all orbital angular momenta was later found \cite{batishchev2007siegert}. The method was demonstrated to work well for obtaining the bound states and scattering properties of a given potential, but the use of Siegert states in self-consistent field calculations has to our knowledge not yet been demonstrated.

In this work, we show that Siegert states are indeed well suited to the solution of average atom models using the numerical scheme laid out in references \cite{tolstikhin97siegert,tolstikhin1998siegert,batishchev2007siegert}.  We demonstrate that the method works over a wide range of density and temperature, and a range of materials, and is numerically efficient.  We start by reviewing the main equations and results of the Siegert state (SS) formalism and how they can be used in average atom models.  Then interesting examples of how pressure ionization gives rise to anti-bound states and resonances in the continuum are presented.  We compare pressures from the model to another state-of-the-art average atom implementation and finally demonstrate convergence with respect to the number of basis functions.

While we restrict ourselves here to an application to average atoms, Siegert states could be much more widely used in the plasma community.  Some applications of interest include evaluation of the electronic structure in calculating the optical or transport properties of plasmas \cite{fontes15, faussurier18, white2022charge, iglesias1996updated, son14, bar1989super, souza2014predictions, hansen2007hybrid, starrett13, perrot1995equation, shaffer17}.

\section{Siegert States for an Average Atom Model}
\subsection{Siegert States}
We start with the radial Schr\"odinger equation in Hartree atomic units
\begin{equation}
\label{eq_se}
    \frac{d^2P_l}{dr^2} + 2 
    \left( 
    \epsilon - V^{eff}(r)-\frac{l(l+1)}{2r^2}
    \right) P_l = 0
\end{equation}
where the radial solution $P_l(r;\epsilon)$ depends on the orbital angular momentum $l$, electron energy $\epsilon$ and radial distance from the nucleus $r$.  $V^{eff}(r)$ is the effective one-electron potential, and
\begin{equation}
    V^{eff}(r) = 0 \mbox{    for    } r\ge R
    \label{eq_veff}
\end{equation}
For the average atom model, $R$ is the ion-sphere radius.  

The physical states are regular at the origin, and behave asymptotically as\footnote{We use an overbar to distinguish the physical states from the Siegert states.}
\begin{equation}
    \begin{split}
        \lim_{r\to\infty} \bar{P}_l(r;\epsilon) = e^{-ikr} - (-\imath)^l S_l(k) e^{ikr}
    \end{split}
\end{equation}
where $k=\sqrt{(2\epsilon)}$ is the complex momentum, and $S_l(k)$ is the S-matrix.  Poles of the S-matrix on the positive imaginary $k$ axis correspond the the bound states.  There are other poles, however, lying in the negative imaginary $k$ plane.  These, as we shall see, correspond to unphysical `states' known as anti-bound, resonant and anti-resonant \cite{belchev11flow,chilcott21experimental, newton2013scattering}.  To find the poles we search for solutions that behave as outgoing waves only
\begin{equation}
    \begin{split}
        \lim_{r\to\infty} P_l(r;\epsilon) = e^{ikr}
    \end{split}
\end{equation}
This is the is the Siegert boundary condition \cite{siegert1939derivation}, alternatively written as
\begin{equation}
    \begin{split}
    \left.
\left(
\frac{d}{dr} -\imath k
\right)P_l(r;\epsilon)\right|_{r\to\infty}
=0
\end{split}
\end{equation}
The eigenstates which satisfy this boundary condition, and are regular at the origin, are known as Siegert (or Gamow-Siegert \cite{rosas2008primer, gamow1928quantentheorie}) states \cite{siegert1939derivation}.

To solve for these states in a robust and efficient way proved difficult in the decades after Siegert's paper.  Quite recently however, an efficient and robust method has been found \cite{tolstikhin97siegert,tolstikhin1998siegert, batishchev2007siegert}.  The basic idea is that for a limited range effective potential $V^{eff}(r)$, satisfying the restriction (\ref{eq_veff}), the solutions can be expanded in a finite basis, reducing the problem to an algebraic form that includes the boundary conditions.  The resulting matrix equation is non-linear in $k$ and is difficult to solve directly.  This is linearized by increasing the dimension of the Hilbert space so that it can be solved by standard methods.  Instead of the original $N$ dimensional space (for $N$ basis functions), the dimension increases to $2N+l$, meaning that for a basis of size $N$, there are $2N+l$ eigenstates $P_{n,l}(r)$ with eigenvalues $k_n$.  

The Siegert states $P_{n, l}(r)$ (we drop the explicit $\epsilon$ dependence for notational simplicity) are normalized according to
\begin{equation}
    \begin{split}
        \int_0^R P_{n,l}(r) P_{m,l}(r) dr +\imath \frac{P_{n,l}(R) P_{m,l}(R) }{k_n+k_m}\\
        \times\left[ 1 + \sum_{p=1}^l \frac{z_{lp}}{(\imath k_n R + z_{lp}) (\imath k_m R + z_{lp})}  \right] = \delta_{nm}
    \end{split}
     \label{eq_norm}
\end{equation}
where $z_{lp}$ are the zeros of the reverse Bessel polynomial \cite{batishchev2007siegert}.  If $P_{m,l}$ and $P_{n,l}$ are bound states of an isolated atom, then this recovers the usual bound state normalization condition \cite{blenski95}
\begin{equation}
    \begin{split}
        \int_0^\infty P_{n,l}(r) P_{m,l}(r) dr = \delta_{nm}
    \end{split}
    \label{eq_bnorm}
\end{equation}
It is worth noting that equations (\ref{eq_norm}) and (\ref{eq_bnorm}) are {\it not} missing a complex conjugate on one of the Siegert states; this is a hallmark of Siegert states.

Some general properties of the Siegert states can be stated.  If $k_n$ is an eigenvalue, so is $-k_n^*$.  If $\Re{k_n} \ne 0$, the complex conjugate pair correspond to a resonant ($\Re{k_n}>0$) and anti-resonant ($\Re{k_n} < 0$) pair.  In this case the resonant and anti-resonant eigenfunctions are complex conjugate.  If $\Re{k_n} = 0$, then for $\Im{k_n} > 0$, the state is a physical bound state, and otherwise ($\Im{K_n}<0$) it is an anti-bound state.  The corresponding eigenfunctions are either purely real or purely imaginary.

Crucially, the physical scattering states $\bar{P}_l(r,\epsilon)$ can be expressed as a sum of Siegert states
\begin{equation}
    \begin{split}
        \bar{P}_l(r) = -\imath k \frac{G_l(r,R;k)}{e_l(kr)}
        \label{eq_pss}
    \end{split}
\end{equation}
where $e_l$ is a special function closely related to the Hankel function (see equation (A8) of reference \cite{batishchev2007siegert}), and the partial-wave Green's function is
\begin{equation}
    \begin{split}
        G_l(r,r';k) = \sum_{n=1}^{2N+l} \frac{P_{n,l}(r) P_{n,l}(r')}{k_n(k-k_n)}
        \label{eq_gfl}
    \end{split}
\end{equation}
The ability to represent the entire continuum of scattering states from a discrete set of Siegert states is surprising and very powerful.
It is this property of Siegert states that makes them appealing to use in an average atom model.

\subsection{Average Atom Model}
Here we give a brief summary of the average atom model and show how the Siegert states are relevant to it.  We consider a nucleus of charge $Z$ at the center of a sphere whose volume is the average volume per atom in the material, $V$, with radius $R$. The sphere is charge neutral, and the electron density inside the sphere is given by the Mermin-Kohn-Sham finite temperature density functional theory \cite{mermin65, kohn1965self, hohenberg}
\begin{equation}
    \begin{split}
        n_e(r) = & \sum_{n,l \in Bound}  f(\epsilon_n,\mu) \frac{2(2l+1)}{4\pi} \frac{\bar{P}_l(r;\epsilon_n)^2}{r^2} \\
                 & + \int_0^\infty d\epsilon f(\epsilon,\mu) \sum_{l=0}^{\infty}\frac{2(2l+1)}{4\pi} \frac{\bar{P}_l(r;\epsilon)^2}{r^2}         
                 \label{eq_ne}
    \end{split}
\end{equation}
where $f(\epsilon,\mu)$ is the Fermi-Dirac occupation factor and $\mu$ is the chemical potential determined by enforcing $Z = 4\pi\int_0^R r^2 n_e(r) dr$.  This expression is arrived at by using the spherical symmetry of the average atom.  By solving the Poisson equation and taking into account the exchange and correlation contribution, the effective potential is
\begin{equation}
    \begin{split}
        V^{eff}(r) =&-\frac{Z}{r} +  \frac{4\pi}{r}\int_0^r\,dr^\prime \,{r^\prime}^2 n_e(r^\prime)\\
         & + 4\pi \int_r^R \,dr^\prime \,{r^\prime} n_e(r^\prime) \\
         & + V^{xc}[n_e(r)] - V^{xc}[n_e(R)]
         \label{eq_veff2}
    \end{split}
\end{equation}
Equations (\ref{eq_se}), (\ref{eq_ne}), and (\ref{eq_veff2}) are solved self-consistently until converged.

However, the electron density can also be found from the Green's function
\begin{equation}
    \begin{split}
        n_e(r) = -\frac{1}{\pi} \Im{ \int_{-\infty}^{\infty} d\epsilon f(\epsilon,\mu) \sum_{l=0}^{\infty} \frac{2(2l+1)}{4\pi} \frac{G_l(r,r;k)}{r^2} }
    \end{split}
\end{equation}
Using equation (\ref{eq_gfl}), it is clear that the electron density can be evaluated directly from the Siegert states without ever calculating the physical scattering states, equation (\ref{eq_pss}).  As pointed out in reference \cite{starrett15}, is it numerically convenient to use Cauchy's integral theorem to rewrite this is
\begin{equation}
    \begin{split}
        n_e(r) = & \frac{1}{\pi} \Im{ \int_C d\epsilon f(\epsilon,\mu) \sum_{l=0}^{\infty} \frac{2(2l+1)}{4\pi} \frac{G_l(r,r;k)}{r^2} }\\
                 & + 2 T \Re{\sum_{j=1}^{N_{mat}} \sum_l \frac{2(2l+1)}{4\pi} \frac{G_l(r,r;\tilde k_j)}{r^2} }
    \end{split}
    \label{eq_ne2}
\end{equation}
where $C$ is a convenient contour, and $N_{mat}$ is the number of Matsubara poles inside the contour.  The Matsubara energies are
\begin{equation}
    \begin{split}
        \tilde\epsilon_j = \mu + \imath \pi (2j-1) T
    \end{split}
\end{equation}
where $T$ is the temperature.

It is interesting to note that if the contour is chosen to be an infinite semi-circle in the upper half energy plane, then the density is simply
\begin{equation}
    \begin{split}
        n_e(r) = 2 T \Re{\sum_{i=1}^{\infty} \sum_l \frac{2(2l+1)}{4\pi} \frac{G_l(r,r;k_i)}{r^2} }
    \end{split}
\end{equation}
and no numerical integration is required.  This is usually not a practical formula due to the difficulty in evaluating the Green's function for the large imaginary arguments that are required for convergence \cite{watrous99green}.  However, the simple dependence of the Green's function on energy (equation (\ref{eq_gfl})) solves this problem.  Nevertheless, for our calculations we have used the formula (\ref{eq_ne2}) with a contour similar to that described in reference \cite{starrett15}.

For later use, we note that the density of states (DOS) is given in terms of the Green's function by
\begin{equation}
    \begin{split}
        \chi(k) = -\frac{1}{\pi} \Im{ \int_{0}^{R} dr \sum_{l=0}^{\infty} 2(2l+1) G_l(r,r;k)}
    \end{split}
\end{equation}
and the partial DOS we define as
\begin{equation}
    \begin{split}
        \chi_l(k) = -\frac{1}{\pi} \Im{ \int_{0}^{R} dr 2(2l+1) G_l(r,r;k)}
    \end{split}
\end{equation}
evaluated using Eq.~\eqref{eq_gfl}.

\section{Results}

\begin{figure}
\begin{center}
\includegraphics[scale=1]{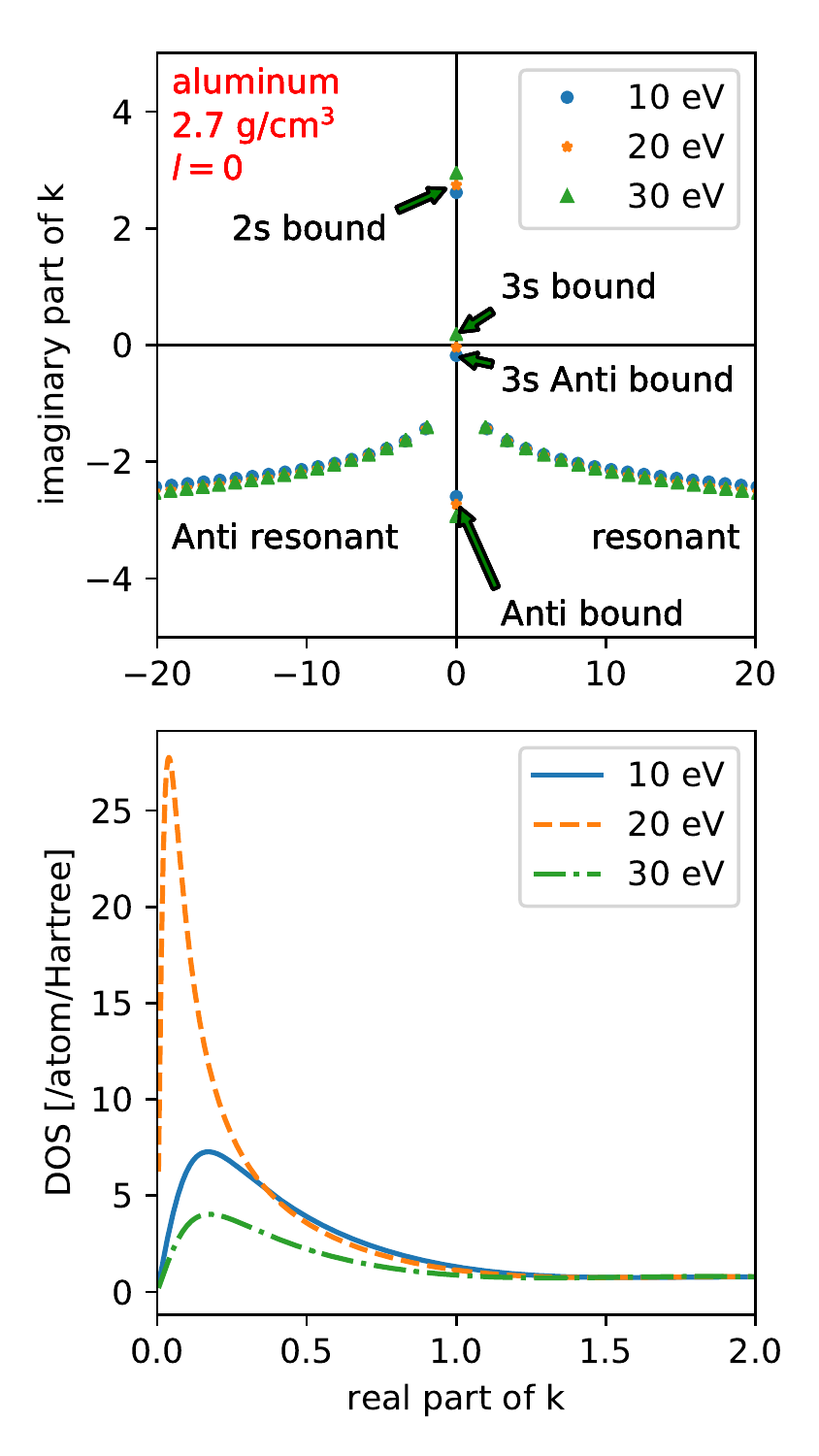} 
\end{center}
\caption{Siegert eigenvalues and density of states (DOS) for aluminum at solid density and temperatures for 10, 20 and 30 eV ($l=0$ only).  Note that the resonant and anti-resonant eigenvalues are nearly on top of each other for the three temperatures.}
\label{fig_eigenv}
\end{figure}

\begin{figure}
\begin{center}
\includegraphics[scale=1]{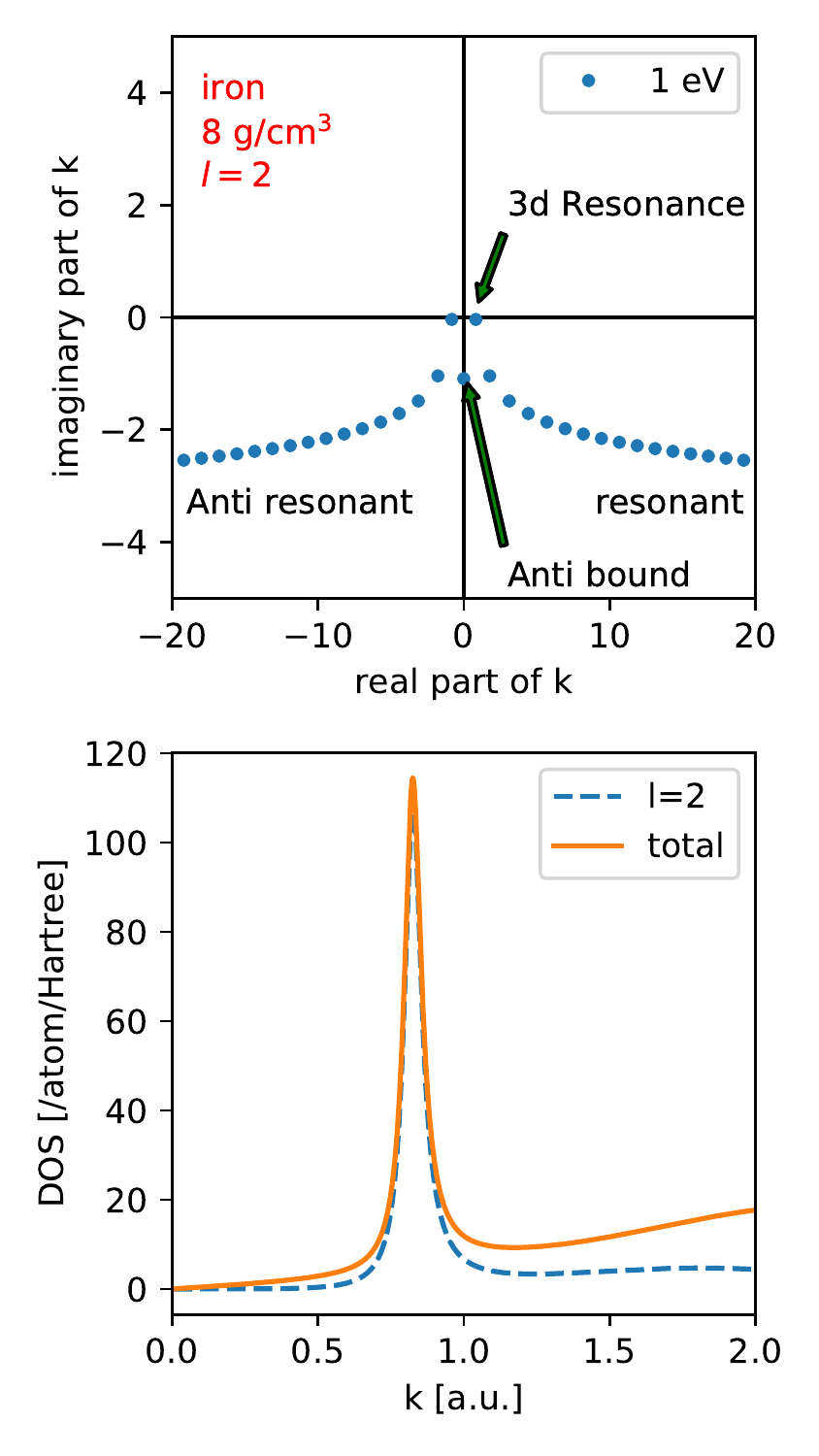} 
\end{center}
\caption{Siegert states and density of states (DOS) for iron at 1 eV temperature and 8 g/cm$^3$.   We show only the eigenvalues for $l=2$ states, where a resonant state lies close to the real energy axis, leading to a resonance in the DOS.}
\label{fig_dos}
\end{figure}

Figure \ref{fig_eigenv} shows an example of Siegert eigenvalues for an aluminum plasma at 2.7 g/cm$^3$ and the corresponding density of states (DOS)  for temperatures of 10, 20 and 30 eV.  We show the eigenvalues in a restricted range of momentum space to highlight certain features, and show the $l=0$ eigenvalues only.  For all three temperatures the $2s$ bound state is apparent on the positive imaginary axis.  Further, in the lower half-plane we see resonant and anti-resonant pairs that are only weakly affected by temperature.  

As temperature is increased, the average energy of the electrons increases and the electrons therefore screen the nucleus more weakly, leading to a deepening of the bound eigenvalues.  This is most clearly seen in the 2$s$ bound state increasing in $\Im k$ with increasing temperature.  Likewise, we see that at 30 eV, the 3$s$ state is bound, but that at lower temperatures, it is anti-bound.  In this case, the nuclear potential is more strongly screened at lower temperature, and the shorter-ranged potential no longer supports a bound 3$s$ state.  

An anti-bound state is distinct from the more commonly recognized resonance state, which occurs for $l\ge 0$ (see later).  Figure \ref{fig_eigenv} shows the effect the anti-bound state has on the DOS.  Going from 10 to 20 eV we see that the appearance of the anti-bound state near the real $k$ axis causes a large peak in the DOS, which quickly reduces as the 3$s$ anti-bound moves further into the negative $k$ plane.
\begin{figure}
\begin{center}
\includegraphics[scale=1]{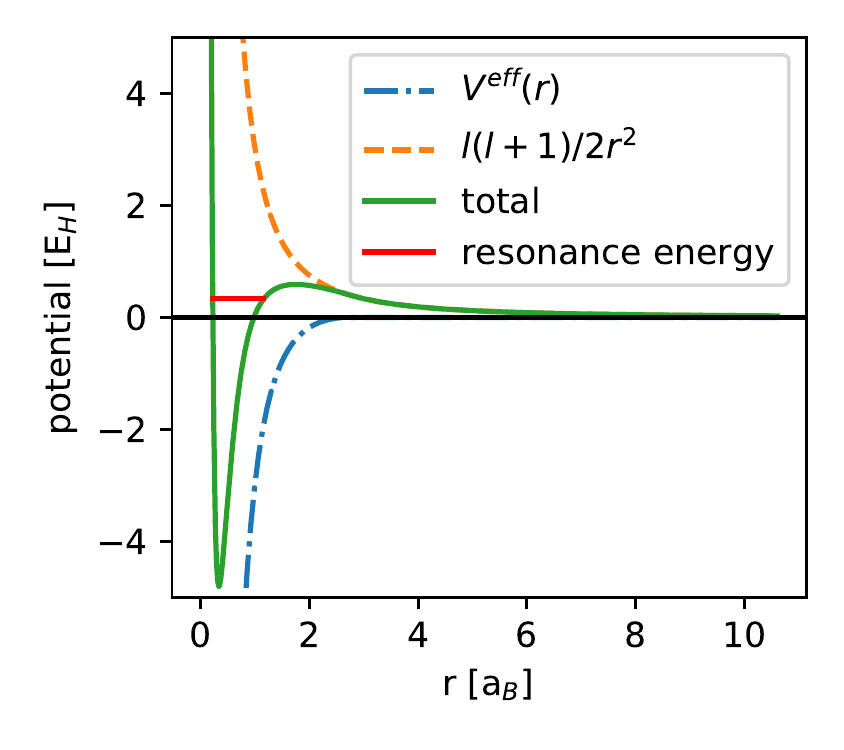} 
\end{center}
\caption{Effective potential $V^{eff}(r)$, centripetal term $-l(l+1)/2r^2$ and total potential for $l=2$, for iron at 8 g/cm$^3$ and 1 eV temperature.  The energy of the $3d$ resonance state ($\epsilon = \Re{k_{n,l}}^2 / 2$) is marked on the plot showing how, through quantum tunneling, the resonance state can decay.}
\label{fig_veff}
\end{figure}

\begin{figure}
\begin{center}
\includegraphics[scale=1]{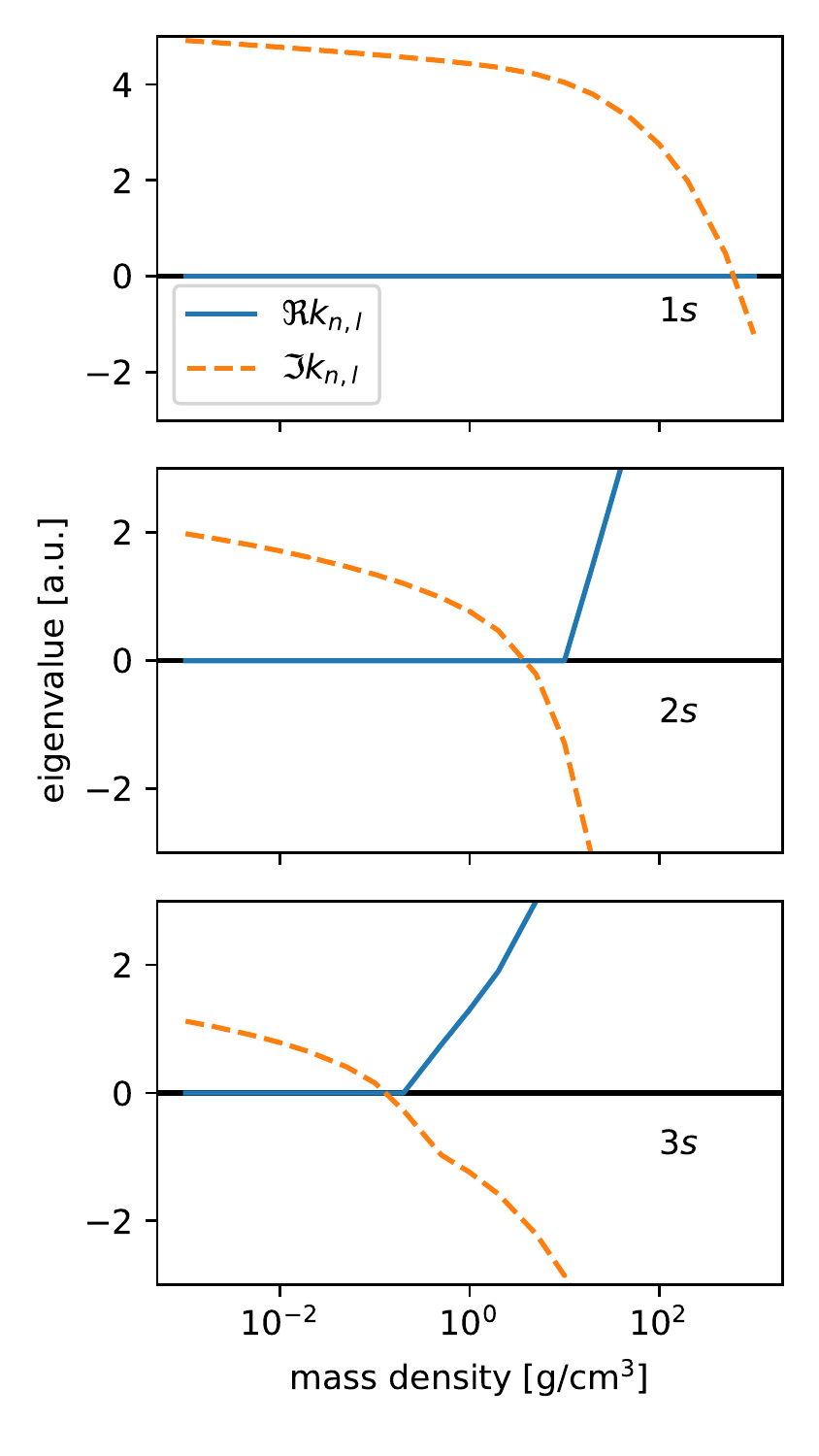} 
\end{center}
\caption{Real and imaginary parts of the Siegert eigenvalue $k_{n,l}$ for carbon at a temperature of  10 eV as a function of mass density.  The top panel shows the evolution of the $1s$ eigenvalue, the middle panel shows the $2s$ and the bottom the $3s$.}
\label{fig_c_s}
\end{figure}

\begin{figure}
\begin{center}
\includegraphics[scale=1]{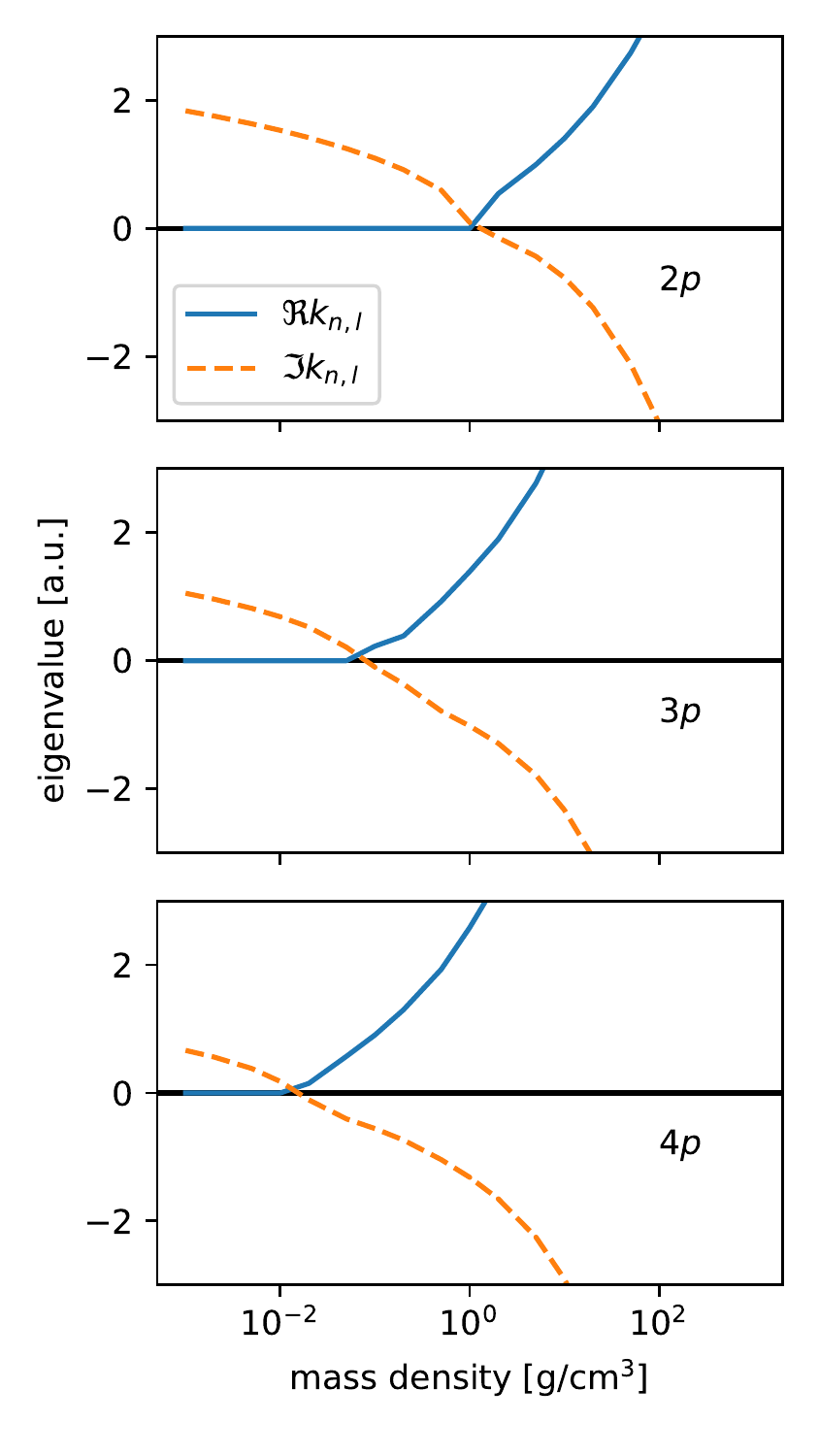} 
\end{center}
\caption{Real and imaginary parts of the Siegert eigenvalue $k_{n,l}$ for carbon at a temperature of  10 eV as a function of mass density.  The top panel shows the evolution of the $2p$ eigenvalue, the middle panel shows the $3p$ and the bottom the $4p$.}
\label{fig_c_p}
\end{figure}
\begin{figure}
\begin{center}
\includegraphics[scale=1]{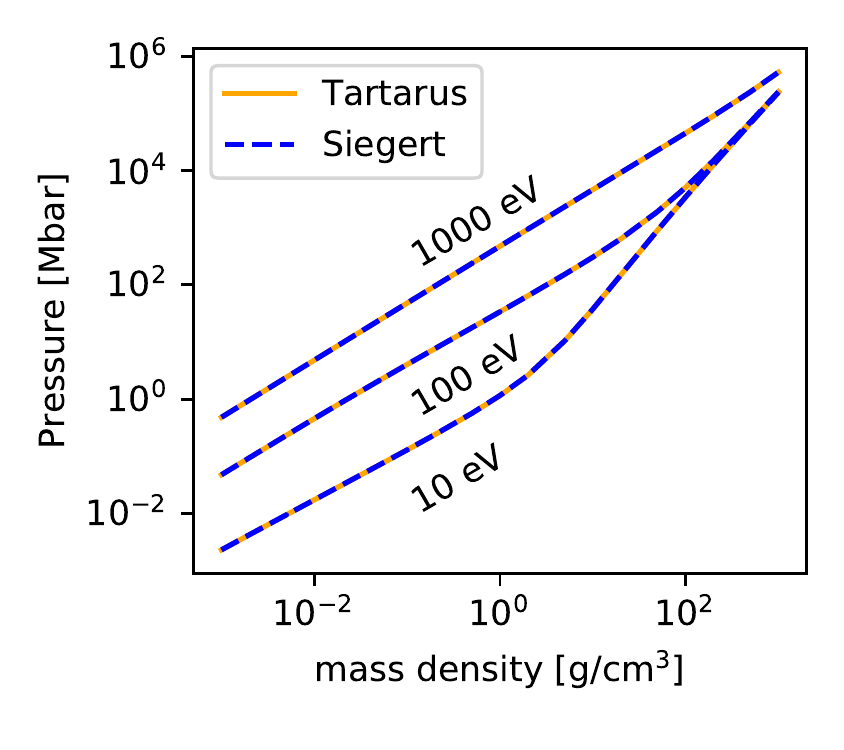} 
\end{center}
\caption{Excess pressure of carbon plasmas at 10, 100 and 1000 eV temperature.  We compare calculations with the Siegert method to results from the 
\texttt{Tartarus} average atom code.  The results of the two methods are on top of each other in the plot, indicating agreement of the methods.}
\label{fig_pexc}
\end{figure}

We can understand the behavior of the DOS due to an anti-bound state by defining the DOS due to one Siegert state as
\begin{equation}
    \begin{split}
        \chi_{n,l}(k) = -\frac{2(2l+1)}{\pi} \Im{ \frac{1}{k_n(k-k_n)} \int_{0}^{R} dr P_{n,l}(r)^2 }
    \end{split}
\end{equation}
For bound (B) or anti-bound (AB) states the the integral term is real, and the eigenvalue is purely imaginary.  Let $k_n = \imath y_n$, where $y_n$ is real, then
\begin{equation}
    \begin{split}
        \chi_{n,l}^{B/AB}(k) = \frac{2(2l+1)}{\pi} \int_{0}^{R} dr P_{n,l}(r)^2 \frac{k}{y_n(y_n^2 + k^2)}
    \end{split}
\end{equation}
We can see that the maximum of the DOS due to an anti-bound state occurs at $k=y_n$, with the maximum value being proportional to $1/(2y_n^2)$.  Therefore, for small $y_n$, the DOS becomes strongly peaked at $k=y_n$.

In figure \ref{fig_dos} the eigenvalues for iron at 8 g/cm$^3$ and 1 eV temperature are shown for $l=2$.  For the range of $k$ shown there are no bound states.  There is one anti-bound, and the rest are resonant or anti-resonant states.  Of particular interest is the resonant state that lies near the positive real axis; this state is the 3$d$ resonance state.  We can see this by looking at the DOS in the bottom panel of the figure.  There we show the total DOS and that due to only the $l=2$ states.  The large peak observed is known as a shape resonance, and physically is due to a potential well created by the effective potential $V^{eff}(r)$ and the centrifugal term $l(l+1)/2r^2$.

 Figure \ref{fig_veff} shows the resonance state with real part of energy $\epsilon = \Re{k_{n,l}}^2 / 2$ is trapped in a potential well created by the sum of $V^{eff}(r)$, which is purely attractive, and the centrifugal term $l(l+1)/2r^2$ which is repulsive.  Because the barrier of the potential well is finite ranged, the lifetime of the resonance state is finite (in contrast to bound states), and an electron in that state can tunnel out.  This quasi-bound character of the resonance state is what gives rise to the characteristic shape resonance feature in the DOS.  In contrast, anti-bound states are not trapped in a potential well, so only those that are close in energy to being trapped lead to such features in the DOS.

One conceptual advantage of Siegert states is that one can directly study the evolution of bound states as a function of density or temperature, even after they ionize. This is because upon ionization, bound states simply transform either into anti-bound or resonant states.  Figures \ref{fig_c_s} and \ref{fig_c_p} show the evolution of selected bound states for a 10 eV carbon plasma over six orders of magnitude in density.  For low densities, the eigenvalue $k_{n,l}$ lies on the imaginary axis.  The more deeply bound the state is at the lowest density, the greater increase in density is needed to pressure ionize it.  Pressure ionization occurs when the imaginary part of $k_{n,l}$ crosses the real axis and becomes negative.  For $p$ states (figure \ref{fig_c_p}) this crossing is immediately accompanied by the real part of $k_{n,l}$ becoming non-zero and positive (we show only the resonant state eigenvalue, the anti-resonant state eigenvalue is just its complex conjugate).  For $s$ states, figure \ref{fig_c_s}, the crossing of the real axis is not accompanied by an immediate increase in the real part.  There is a delay in which the real part remains zero - the bound state becomes and anti-bound state before splitting into a resonant/anti-resonant pair.

The physical reason for this difference in behaviors between $s$-states and all other orbital angular momenta states is the centrifugal barrier.  The mathematical reason is explored in reference \cite{batishchev2007siegert}.  The authors show that for $l \ge 1$ there is a forbidden region of $k$-space (a `dead-zone' in their language) where eigenvalues cannot occur. This dead zone touches the origin and lies in the lower half-plane. As a result, when a bound state with $l\ge1$ ionizes, it cannot continue down the imaginary $k$ axis and must instead split into a resonant/anti-resonant pair. 
%This dead-zone is defined by the domain of negativity of the function
%\begin{equation}
%    \begin{split}
%        w_l(z) = 1 + \sum_{p=1}^l \frac{Z_{lp}}{(z-z_{lp})(z^*-z_{lp})}
%    \end{split}
%\end{equation}

As a demonstration of the usefulness of the method for equation of state calculations, in figure \ref{fig_pexc} we show the excess pressure of carbon plasma over a wide range of temperatures and densities.  The excess pressure is defined as the total pressure minus an ideal ion contribution $k_B T/V$.  In the figure we compare to results from the \texttt{Tartarus} average atom code, which implements the same model using a more conventional finite difference scheme \cite{starrett19}.  No appreciable difference is observed between the two methods, and the computational cost is similar for each, typically 1-2 minutes per density and temperature.  It is difficult to fairly make a more precise comparison, as wall-times depend on convergence parameters such as radial grid size, number of quadrature points for the energy integral, etc., that we have not systematically optimized in either case. Nevertheless, an advantage that the finite-basis Siegert approach has over finite-difference integration schemes is that a sparser radial grid can be used to achieve comparable accuracy, which lowers the memory footprint of the code.  For reference, we use 300 radial grid points for the spectral Siegert method versus 3000 for the finite-difference scheme in \texttt{Tartarus}.

A further advantage of the Siegert method is that only one type of solver is need for all states.  For the traditional approach, one might use a inward and outward integration scheme combined with a search technique for finding the bound state solutions, an outward integration method and resonance tracking scheme for the continuum electron, or, as in \texttt{Tartarus}, a different method again to find the Green's function.  This range of techniques leads to algorithmic complexity.  In the Siegert method, one matrix diagonalization, for which standard libraries exist, gives the entire eigenspectrum.
\begin{figure}[H]
\begin{center}
\includegraphics[scale=1]{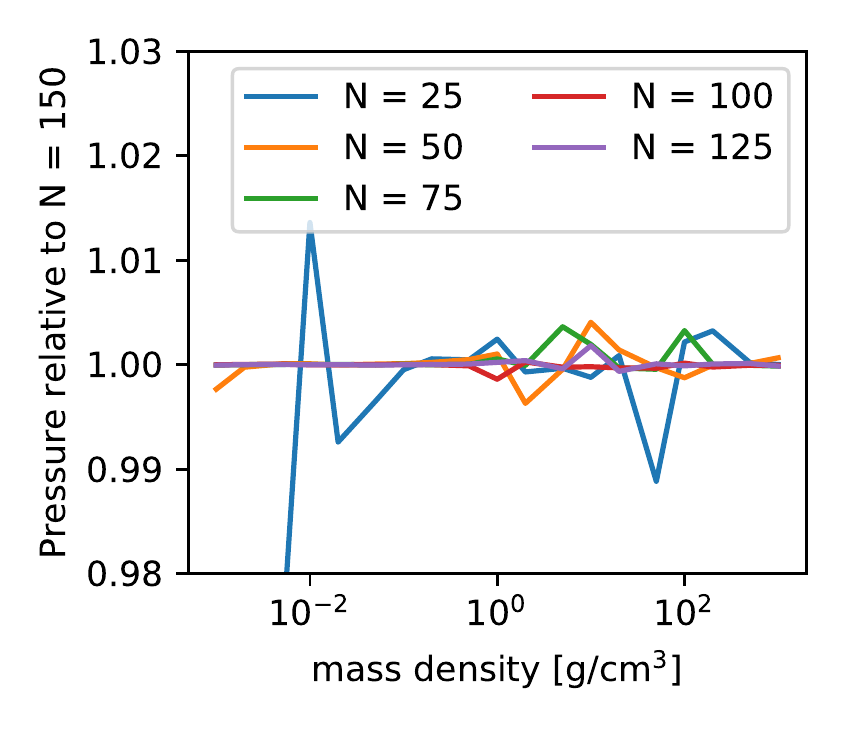} 
\end{center}
\caption{Convergence of the excess pressure with respect to the size of the basis $N$ for a carbon plasma at 10 eV temperature.  Errors in the excess pressure are calculated relative to $N=150$.  }
\label{fig_pexr}
\end{figure}

\begin{figure}
\begin{center}
\includegraphics[scale=1]{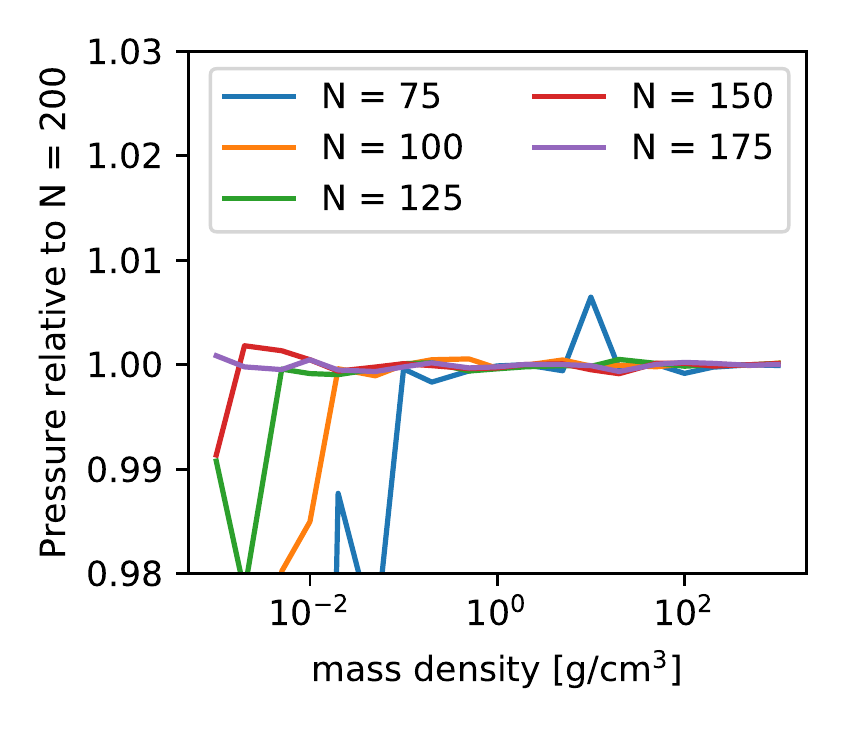}
\end{center}
\caption{As for figure \ref{fig_pexr} but for copper plasmas at 10 eV temperature, and errors have been calculated relative to $N=200$. }
\label{fig_pex_cu}
\end{figure}

In figures \ref{fig_pexr} and \ref{fig_pex_cu} we show pressures from carbon and copper plasmas as a function of the size of the basis $N$.  In each case we show the ratio of the pressure to the converged result.  Increasing the size of the basis has two distinct effects.  On the one hand it allows more complicated radial functions to be accurately represented.   On the other hand, it increases the maximum $|k_{n,l}|$ that is accurately represented\footnote{Note that we have used the usual average atom acceleration scheme in which it is the difference of the continuum electron density and that of truly free electrons that is important.  See section 3.3 of reference \cite{starrett19}.}.  It is remarkable that for carbon, as few as 50 basis functions give errors less than 1\% for a wide range of mass densities and temperatures, and errors very much less than 1\% are found for $N=125$.  For reference we used $N=150$ for all the results presented up to this point.

For copper, figure \ref{fig_pex_cu}, a larger basis is needed to maintain a similar accuracy.  This is expected, as more highly charged nuclei require accurate representation of higher energy states.  This is also why the figure shows that a larger basis is needed at lower densities.  Lower densities for the same temperature correspond to a less degenerate system, meaning that the tail of the Fermi-Dirac distribution will be longer.  Presumably this does not show up for carbon due to the error being dominated by inadequate representation of the radial functions for $N=25$.

\section{Conclusions}
Siegert states have been shown to be a viable and advantageous method for use in average atom models.  It has been demonstrated that the Siegert states can be used to accurately construct the Green's function and in turn the electron density needed to perform self-consistent field calculations.  We investigated two pressure ionization scenarios; that in which an $s$-wave state is ionized to become an anti-bound and subsequently a resonant and anti-resonant pair, and that for non-zero orbital angular momenta, in which pressure ionized bound states directly turn into resonant and anti-resonant states.  We show how both scenarios lead to spikes in the density of states.  It was also demonstrated that the method leads to accurate pressures when compared to other state-of-the-art average atom methods.

The Siegert state approach results in a completely discrete set of eigenstates that can be used to represent the Green's function and the scattering states without approximation.  The advantage of a discrete set of states could be further exploited through their use in perturbation theory \cite{more1985pressure} or expanding time-dependent wave packets \cite{santra2005siegert}, among other applications requiring a complete set of states where it would be impractical to use the usual continuum wave functions.

\section*{Acknowledgments}
We thank D. Saumon for useful discussions on the early part of this work. This work was support by LANL’s ASC PEM Atomic Physics Project. LANL is operated by Triad National Security, LLC, for the National Nuclear Security Administration of the U.S. Department of Energy under Contract No.~89233218NCA000001. N.R.S. acknowledges support by the Department of Energy National Nuclear Security Administration under Award Number DE-NA0003856, and the New York State Energy Research and Development Authority.

\bibliographystyle{unsrt}
\bibliography{phys_bib}

\end{document}